\documentclass[aps,preprint, floatfix]{revtex4}

\usepackage{graphicx,color}
\usepackage{psfrag}
\usepackage{amssymb}
\usepackage{eso-pic}
\usepackage[dvips,letterpaper]{geometry} 
\usepackage{amsmath}

\begin{document}

%
%
%
%
\def\oti{{\otimes}}
\def\lb{ \left[ }
\def\rb{ \right]  }
\def\tilde{\widetilde}
\def\bar{\overline}
\def\hat{\widehat}
\def\*{\star}
\def\[{\left[}
\def\]{\right]}
\def\({\left(}		\def\BL{\Bigr(}
\def\){\right)}		\def\BR{\Bigr)}
	\def\BBL{\lb}
	\def\BBR{\rb}
%
%
\def\zb{{\bar{z} }}
\def\zbar{{\bar{z} }}
\def\frac#1#2{{#1 \over #2}}
\def\inv#1{{1 \over #1}}
\def\half{{1 \over 2}}
\def\d{\partial}
\def\der#1{{\partial \over \partial #1}}
\def\dd#1#2{{\partial #1 \over \partial #2}}
\def\vev#1{\langle #1 \rangle}
\def\ket#1{ | #1 \rangle}
\def\rvac{\hbox{$\vert 0\rangle$}}
\def\lvac{\hbox{$\langle 0 \vert $}}
\def\2pi{\hbox{$2\pi i$}}
\def\e#1{{\rm e}^{^{\textstyle #1}}}
\def\grad#1{\,\nabla\!_{{#1}}\,}
\def\dsl{\raise.15ex\hbox{/}\kern-.57em\partial}
\def\Dsl{\,\raise.15ex\hbox{/}\mkern-.13.5mu D}
%
%
\def\ga{\gamma}		\def\Ga{\Gamma}
\def\be{\beta}
\def\al{\alpha}
\def\ep{\epsilon}
\def\vep{\varepsilon}
\def\la{\lambda}	\def\La{\Lambda}
\def\de{\delta}		\def\De{\Delta}
\def\om{\omega}		\def\Om{\Omega}
\def\sig{\sigma}	\def\Sig{\Sigma}
\def\vphi{\varphi}

%
%
\def\CA{{\cal A}}	\def\CB{{\cal B}}	\def\CC{{\cal C}}
\def\CD{{\cal D}}	\def\CE{{\cal E}}	\def\CF{{\cal F}}
\def\CG{{\cal G}}	\def\CH{{\cal H}}	\def\CI{{\cal J}}
\def\CJ{{\cal J}}	\def\CK{{\cal K}}	\def\CL{{\cal L}}
\def\CM{{\cal M}}	\def\CN{{\cal N}}	\def\CO{{\cal O}}
\def\CP{{\cal P}}	\def\CQ{{\cal Q}}	\def\CR{{\cal R}}
\def\CS{{\cal S}}	\def\CT{{\cal T}}	\def\CU{{\cal U}}
\def\CV{{\cal V}}	\def\CW{{\cal W}}	\def\CX{{\cal X}}
\def\CY{{\cal Y}}	\def\CZ{{\cal Z}}

\def\rvac{\hbox{$\vert 0\rangle$}}
\def\lvac{\hbox{$\langle 0 \vert $}}
\def\comm#1#2{ \BBL\ #1\ ,\ #2 \BBR }
\def\2pi{\hbox{$2\pi i$}}
\def\e#1{{\rm e}^{^{\textstyle #1}}}
\def\grad#1{\,\nabla\!_{{#1}}\,}
\def\dsl{\raise.15ex\hbox{/}\kern-.57em\partial}
\def\Dsl{\,\raise.15ex\hbox{/}\mkern-.13.5mu D}
%
%
%
\font\numbers=cmss12
\font\upright=cmu10 scaled\magstep1
\def\stroke{\vrule height8pt width0.4pt depth-0.1pt}
\def\topfleck{\vrule height8pt width0.5pt depth-5.9pt}
\def\botfleck{\vrule height2pt width0.5pt depth0.1pt}
\def\Zmath{\vcenter{\hbox{\numbers\rlap{\rlap{Z}\kern
0.8pt\topfleck}\kern 2.2pt
                   \rlap Z\kern 6pt\botfleck\kern 1pt}}}
\def\Qmath{\vcenter{\hbox{\upright\rlap{\rlap{Q}\kern
                   3.8pt\stroke}\phantom{Q}}}}
\def\Nmath{\vcenter{\hbox{\upright\rlap{I}\kern 1.7pt N}}}
\def\Cmath{\vcenter{\hbox{\upright\rlap{\rlap{C}\kern
                   3.8pt\stroke}\phantom{C}}}}
\def\Rmath{\vcenter{\hbox{\upright\rlap{I}\kern 1.7pt R}}}
\def\Z{\ifmmode\Zmath\else$\Zmath$\fi}
\def\Q{\ifmmode\Qmath\else$\Qmath$\fi}
\def\N{\ifmmode\Nmath\else$\Nmath$\fi}
\def\C{\ifmmode\Cmath\else$\Cmath$\fi}
\def\R{\ifmmode\Rmath\else$\Rmath$\fi}

\def\barray{\begin{eqnarray}}
\def\earray{\end{eqnarray}}
\def\beq{\begin{equation}}
\def\eeq{\end{equation}}

\def\n{\noindent}

\def\Tr{\rm Tr} 
\def\xvec{{\bf x}}
\def\kvec{{\bf k}}
\def\kvecp{{\bf k'}}
\def\omk{\om{\kvec}} 
\def\dk#1{\frac{d\kvec_{#1}}{(2\pi)^d}}
\def\2pid{(2\pi)^d}
\def\ket#1{|#1 \rangle}
\def\bra#1{\langle #1 |}
\def\vol{V}
\def\adag{a^\dagger}
\def\rme{{\rm e}}
\def\Im{{\rm Im}}
\def\pvec{{\bf p}}
\def\fermiS{\CS_F}
\def\cdag{c^\dagger}
\def\adag{a^\dagger}
\def\bdag{b^\dagger}
\def\vvec{{\bf v}}
\def\muhat{{\hat{\mu}}}
\def\vac{|0\rangle}
\def\pcut{{\Lambda_c}}
\def\chidot{\dot{\chi}}
\def\gradvec{\vec{\nabla}}
\def\psitilde{\tilde{\Psi}}
\def\psibar{\bar{\psi}}
\def\psidag{\psi^\dagger} 
\def\m{m_*}
\def\up{\uparrow}
\def\down{\downarrow}
\def\Qo{Q^{0}}
\def\vbar{\bar{v}}
\def\ubar{\bar{u}}
\def\smallhalf{{\textstyle \inv{2}}}
\def\smallsqrt{{\textstyle \inv{\sqrt{2}}}}
\def\rvec{{\bf r}}
\def\avec{{\bf a}}
\def\pivec{{\vec{\pi}}}
\def\svec{\vec{s}} 
\def\phivec{\vec{\phi}}
\def\daggerc{{\dagger_c}}
\def\Gfour{G^{(4)}}
\def\dim#1{\lbrack\!\lbrack #1 \rbrack\! \rbrack }
\def\qhat{{\hat{q}}}
\def\ghat{{\hat{g}}}
\def\nvec{{\vec{n}}}
\def\bull{$\bullet$}
\def\ghato{{\hat{g}_0}}
\def\r{r}
\def\deltaq{\delta_q}
\def\gcharge{g_q}
\def\gspin{g_s}
\def\deltas{\delta_s}
\def\gQC{g_{AF}} 
\def\ghatqc{\ghat_{AF}}
\def\xqc{x_{AF}}
\def\mhat{\hat{m}}
\def\xup{x_2}
\def\xdown{x_1}
\def\sigmavec{\vec{\sigma}}
\def\xopt{x_{\rm opt}}
\def\Lambdac{{\Lambda_c}}
\def\angstrom{{{\scriptstyle \circ} \atop A}     }
\def\AA{\leavevmode\setbox0=\hbox{h}\dimen0=\ht0 \advance\dimen0 by-1ex\rlap{
\raise.67\dimen0\hbox{\char'27}}A}
\def\ratio{\gamma}
\def\Phivec{{\vec{\Phi}}}
\def\singlet{\chi^- \chi^+} 
\def\mhat{{\hat{m}}}

\def\Im{{\rm Im}}
\def\Re{{\rm Re}}

\def\xstar{x_*}

\def\sech{{\rm sech}}

\def\Li{{\rm Li}}

\def\dim#1{{\rm dim}[#1]}

\def\ep{\epsilon}

\def\free{\CF}

\def\Fhat{\digamma}

\def\ftilde{\tilde{f}}

\def\muphys{\mu_{\rm phys}}

\def\xitilde{\tilde{\xi}}

\def\CI{\mathcal{I}}

\def\nhat{\hat{n}}

\def\ef{\epsilon_F}

\title{S-matrix approach to quantum  gases in the unitary limit II: 
 the three-dimensional case}
\author{Pye-Ton How and Andr\'e  LeClair}
\affiliation{Newman Laboratory, Cornell University, Ithaca, NY}

\bigskip\bigskip\bigskip\bigskip

\begin{abstract}

A new analytic treatment of  
 three-dimensional homogeneous  Bose and Fermi gases  
in the unitary limit of negative infinite scattering length 
is presented, based on the S-matrix approach to statistical 
mechanics we recently developed.   The unitary limit occurs 
at a fixed point of the renormalization group with dynamical exponent 
$z=2$ where the S-matrix equals $-1$.   For fermions we find
$T_c /T_F \approx 0.1$.   For bosons we present evidence that the
gas does not collapse, but rather has a critical point 
that is
a strongly interacting form of Bose-Einstein condensation.
This bosonic critical point occurs at $n \lambda_T^3 \approx 1.3$
where $n$ is the density and $\lambda_T$ the thermal wavelength,
which is lower than the ideal gas value of $\zeta (3/2) = 2.61$.

\end{abstract}

\maketitle

\section{Introduction}

In the so-called unitary limit of a quantum Bose or Fermi gas,  the
scattering length $a$ diverges.   This occurs at a fixed point of the 
renormalization group,  thus these systems provide interesting examples of 
interacting,  scale-invariant theories with dynamical exponent $z=2$, 
i.e. non-relativistic.  
They can be realized experimentally by tuning the scattering
length to $\pm \infty$ using a Feshbach resonance.
(See for instance \cite{Experiment1,Experiment2} and references
therein.)   They are also thought to occur at the surface of
neutron stars.  These systems  have also attracted much theoretical 
interest\cite{Leggett,Nozieres,Ohashi,Ho,HoMueller,Astrakharchik,Nussinov,Perali,LeeShafer,Wingate,Bulgac,Drummond,Burovski,Nishida,Nikolic}.   
There have even been some proposals to use the AdS/CFT correspondence 
to learn about these models\cite{Son,Maldacena,Herzog,Adams}.

Because of the scale-invariance,  the only length scales in the problem
are based on the density $n^{1/d}$ where $d$ is the spatial dimension, 
and the thermal wavelength $\lambda_T = \sqrt{2\pi/mT}$.   Equivalently,
the only energy scales are the chemical potential $\mu$ and the temperature $T$.  
The problem is challenging since there is no small paramater to expand in
such as $n a^3$.    Any possible critical point must occur at a 
specific value of $x=\mu/T$.   This can be translated into 
universal values for $n_c \lambda_T^3$, or for fermions 
universal values for $T_c/T_F$ where $\ep_F = k_B T_F$ is the Fermi energy.  
For instance the critical point of an ideal Bose gas is the simplest example,
where $n_c \lambda_T^3 = \zeta (3/2) = 2.61$.

The present work is the sequel to \cite{PyeTon2}, where we used the S-matrix 
based formulation
of the quantum statistical mechanics developed in\cite{LeClairS,PyeTon}.   
This approach is very well-suited to the problem because in the unitary limit
the S-matrix $S=-1$, and kernels in the integral equations simplify.   
In fact, this approach can be used to develop an expansion in $1/a$.  
The main formulas for the 2 and 3 dimensional cases of both bosons and fermions
were presented,  however only the 2-dimensional case was analyzed in detail
in \cite{PyeTon2}.   
Here we analyze the 3-dimensional case.

The models considered  are the simplest 
models of non-relativistic bosons or  fermions with quartic
interactions.    The bosonic model is defined by the action
for a complex scalar field $\phi$. 
\beq
\label{bosonaction}
S =  \int d^3 \xvec dt \(  i \phi^\dagger  \d_t \phi - 
\frac{ |\vec{\nabla} \phi |^2}{2m}  - \frac{g}{4} (\phi^\dagger  \phi)^2 \)
\eeq
For fermions, due to the fermionic statistics, 
one needs at least a 2-component field 
$\psi_{\up , \down} $:
\beq
\label{fermionaction}
S = \int d^3 \xvec dt \(  \sum_{\alpha=\up, \down}  
i \psi^\dagger_\alpha \d_t  \psi_\alpha  - 
\frac{|\vec{\nabla}  \psi_\alpha|^2}{2m}   - \frac{g}{2} 
\psi^\dagger_\up \psi_\up \psi^\dagger_\down \psi_\down \) 
\eeq
In both cases,  positive $g$ corresponds to repulsive interactions.   
The bosonic theory only has a $U(1)$ symmetry.   The fermionic theory
on the other hand has the much larger SO(5) symmetry.   
This is evident from the work\cite{Kapit} which considered a relativistic
version, since  the same arguments apply to a non-relativistic kinetic term. 
This is also clear from the work\cite{Nikolic} which considered 
an $N$-component version with Sp(2N) symmetry,  and noting that
Sp(4) = SO(5).

The interplay between the scattering length, the bound state,  
and the renormalization group fixed point was discussed 
in detail from the point of view of the
S-matrix in \cite{PyeTon2}.   In 3 spatial dimensions the fixed point occurs
at negative coupling $g_*= - 4 \pi^2 /m\Lambda$,  where $\Lambda$ is 
an ultra-violet cut-off.  For $g$  less than $g_*$, there is a bound
state that can Bose-Einstein condense (BEC), and for the fermionic case, 
this is referred to as the BEC side.  As $g_*$ is approached from this side,
the scattering length goes to $+\infty$.   The bound state disappears at
$g_*$.  When $g$ approaches $g_*$ from above,  the scattering length 
goes to $-\infty$, and this is referred to as the BCS side.   In this paper we
work on the BCS side of the cross-over since the bound state does not have
to be incorporated into the thermodynamics.  On this side the interactions
are effectively attractive.

Theoretical studies have mainly focussed on the fermionic case,   
and for the most part at zero temperature,   which is appropriate for a large Fermi energy.
  The bosonic case has been less studied,  since 
a homogeneous 
bosonic gas with attractive interactions is thought to be unstable against
mechanical collapse, and the collapse occurs before any kind of BEC.  
The situation is actually different for harmonically  trapped gases,  where BEC can occur\cite{trapped}.  
However studies of the homogeneous bosonic case were  based on a small, 
negative scattering length\cite{Stoof,MuellerBaym,Thouless,YuLiLee},
and it is not clear that the conclusions reached  there can be extrapolated 
to the unitary limit.    Since the density of
collapse is proportional to $1/a$\cite{MuellerBaym},   extrapolation to infinite scattering length 
suggests that the gas  collapses at zero density,  which seems unphysical,  since the gas could in
principle be stabilized at finite temperature by thermal pressure.   
 One can also point out that  in the van der Waals gas,  the
collapse is stabilized by a finite size of the atoms,  which renders
the compressibility finite.  In the unitary limit, there is nothing to play
such a role.  
  In the sequel we will present evidence that the unitary 
Bose gas undergoes BEC when $n \lambda_T^3 \approx 1.3$. 
 This lower value
is consistent with the attractive interactions.    We also estimate 
the critical exponent describing how the compressibility diverges at the critical point.

In the next section we define the  scaling functions  that determine the free energy
and density,  and derive expressions for the energy and entropy per particle,  specific heat per
particle, and compressibility.     In section III the formulation of the unitary limit in
\cite{PyeTon2}  is summarized.     The two-component fermion case is analyzed in section IV.
Evidence is presented for a critical point with $T_c/T_F \approx 0.1$,  which is consistent with
lattice  Monte Carlo simulations.    Bosons are analyzed in section V,  where we present evidence
for BEC in this strongly interacting gas.     Motivated by  the conjectured lower bound\cite{Kovtun} 
for
the ratio of the viscosity to entropy density  $\eta/s > \hbar/4\pi k_B$ for relativistic systems,
we study this ratio for both the fermionic and bosonic cases in section VI.
   Our results for fermions are
consistent with experiments,  with  $\eta/s >   4.72$  times the conjectured lower bound.
For bosons,  this ratio is minimized at the critical point  where 
$\eta/s > 1.26$ times the bound.

\section{Scaling functions in the unitary limit.}

The scale invariance in the unitary limit implies some
universal scaling forms\cite{Ho}.   In this section we 
define various scaling functions with a meaningful normalization
relative to free particles.    

First consider a single species of bosonic or fermionic particle with
mass $m$ at chemical potential $\mu$ and temperature $T$. 
  The free energy density
has the form
\beq
\label{freeenergy}
\CF =  - \zeta (5/2) \( \frac{mT}{2\pi} \)^{3/2} T  \, c(\mu/T) 
\eeq
where the scaling function $c$ is only a function of $x\equiv \mu/T$.
($\zeta$ is Riemann's zeta function.)   
The combination $\sqrt{mT/2\pi} = 1/\lambda_T$, where 
$\lambda_T$ is the thermal wavelength.   For a single free boson  or fermion:
\beq
\label{cfreelim}
\lim_{\mu/T \to 0} ~~  c_{\rm boson} = 1, ~~~~~~~
 c_{\rm fermion} = 1- \inv{2 \sqrt{2}}
~~~~~~({\rm free ~ particles}).  
\eeq
It is also convenient to define the  scaling function $q$, which is a measure
of the quantum degeneracy, in terms of the density   as follows: 
\beq
\label{nhat}
n \lambda_T^3 = q
\eeq
The two scaling functions $c$ and $q$ are of course related since
$n= - \d \CF / \d \mu$,  which leads to 
\beq
\label{qcprime}
q = \zeta (5/2) c'
\eeq   
where $c'$ is the derivative of $c$ with respect to $x$.   
Henceforth $g'$ will always denote the derivative of $g$ with respect to $x$.
The expressions  for $c$ and $q$ for free theories will be implicit
in the next section.

Also of interest are several energy per particle scaling functions.   
At a renormalization group fixed point, the energy density
is related to the free energy in the same way as for free particles:
\beq
\label{energyden}
\frac{E}{V} = -\frac{3}{2} \CF 
\eeq
where $V$ is the volume.    For a free fermion,   in the zero 
temperature limit,  the energy per particle $E/N \to \frac{3}{5} \mu$.
The Fermi energy is 
\beq
\label{EF}
\ef =\inv{m}  \( 3 \pi^2 n/\sqrt{2} \)^{2/3} 
\eeq
The above definition can  also  be used for bosons.  
Since $\ef = \mu$ in the zero temperature free  fermionic gas,  this leads
us to define the scaling function $\xi$:
\beq
\label{xi}
\xi (x)  =  \frac{5}{3} \frac{E}{N \ef} =  \frac{5 \zeta(5/2)}{3}
\( \frac{6}{\pi} \)^{1/3}      \,       \frac{c}{q^{5/3}}
\eeq
In the limit $T\to 0$,  i.e. $x\to \infty$,   $\xi \to 1$ for a free
fermion.   

A different energy per particle scaling function, $\xitilde$,  is meaningful
as $x\to 0$:
\beq
\label{xitilde}
\frac{E}{N T}  =  \frac{3 (2 \sqrt{2} -1 ) \zeta(5/2) }{2(2 \sqrt{2} -2) 
\zeta(3/2)} 
   \,  \xitilde (x) 
\eeq
With the above normalization $\xitilde = 1$ for a free fermion in the limit
$x\to 0$.   In terms of the above scaling functions:
\beq
\label{xitildesc}
\xitilde (x) =   \frac{ \zeta (3/2)   
( 2 \sqrt{2} -2)}{(2 \sqrt{2} -1)}  \,  \frac{c}{q} 
\eeq

The entropy density is $s = - \d \CF / \d T$,  and the entropy per particle
takes the form
\beq
\label{sn}
\frac{s}{n} =   \zeta (5/2) 
\( \frac{ 5 c/2 - x c'}{q} \) 
\eeq

Next consider the 
 specific heat per particle at constant volume and particle number,
i.e. constant density.   One needs $\d x/ \d T$ at constant density.
Using the fact that $n \propto T^{3/2} q$,  at constant density
$q \propto T^{-3/2}$.   This gives
\beq
\label{CV.1}
T \(  \frac{\d x} {\d T} \)_n =   -\frac{3}{2}  \frac{ q}{q'} 
\eeq
The specific heat per particle is then:
\beq
\label{CV.2}
\frac{C_V}{N} = \inv{N}  \( \frac{\d E}{\d T} \)_{N,V} =
\frac{\zeta(5/2)}{4} \( 15 \frac{c}{q} - 9 \frac{c'}{q'} \) 
\eeq

The isothermal compressibility is defined as 
\beq
\label{comp.1}
\kappa =  - \inv{V} \( \frac{ \d V}{\d p} \)_T 
\eeq
where the pressure $p= - \CF$.   Since $n=N/V$ and $N$ is kept fixed, 
\beq
\label{comp.2}
\kappa =  - n  \( \frac{ \d n^{-1} }{\d p} \)_T =  \inv{n T}  \frac{ q'}{q}  =
\inv{T} \( \frac{mT}{2\pi} \)^{3/2}   \frac{q'}{q^2}   
\eeq

Finally the equation of state can be expressed parametrically 
as follows.  Given $n$ and $T$, one uses eq. (\ref{nhat}) to find
$x$ as a function of $n,T$.   The pressure can then be written as
\beq
\label{eqnstate}  
p =  \(  \frac{\zeta(5/2) c(x(n,T))}{q(x(n,T))} \)  n T
\eeq

In order to compare with numerical simulations and experiments,
it will be useful to plot various quantities as a function of  $q$ or 
$T/T_F$:
\beq
\label{TTF}
\frac{T}{T_F}  =  \(  \frac{4}{3 \sqrt\pi  q} \)^{2/3} 
\eeq

\section{Two-body scattering approximation}

The main features of the two-body scattering approximation developed in
\cite{PyeTon} are the following.   Consider again first a single component
gas.   The filling fractions,  or 
occupation numbers,  are parameterized in terms of a pseudo-energy
$\vep (\kvec )$:
\beq
\label{fill}
f(\kvec )  =  \inv{ e^{\beta \vep (\kvec ) } -s }
\eeq
which determine the density:
\beq
\label{dens}
n = \int \frac{d^3 \kvec}{(2\pi)^3} ~ \inv{ e^{\beta \vep  (\kvec ) } -s }
\eeq
where $s=1,-1$ corresponds to bosons, fermions respectively
and $\beta = 1/T$.    
The consistent summation of 2-body scattering leads to 
an integral equation for the 
 pseudo-energy $\vep (\kvec)$.   It is convenient to define the
quantity:
\beq
\label{ydef}
y (\kvec ) = e^{-\beta (\vep(\kvec) - \omega_\kvec + \mu )}
\eeq
where $\omega_\kvec = \kvec^2 / 2m$.   Then $y$ satisfies the integral
equation 
\beq
\label{yinteq}
y (\kvec ) =  1 + \beta   \int  \frac{d^3 \kvec'}{(2\pi)^3} \, 
G(\kvec - \kvec' )  \frac{y(\kvec' )^{-1}}{e^{\beta \vep (\kvec')} -s}
\eeq
The free energy density is then
\beq
\label{freefoam2}
\CF = -T  \int \frac{d^3 \kvec}{ (2\pi)^3} \[ 
- s  \log ( 1- s e^{-\beta \vep  } ) 
-\inv{2}  \frac{ (1-y^{-1} )}{e^{\beta \vep} -s}   \]  
\eeq

The kernel has the following structure:
\beq
\label{Gstructure}
G  =  - \frac{i}{\CI} \log ( 1 + i \CI \CM)
\eeq
where $\CM$ is the scattering amplitude and $\CI$ represents
the available phase space for two-body scattering.   The argument
of the $\log$ can be identified as the S-matrix function. 
In the unitary limit,  
\beq
\label{CMunit}
\CM =   \frac{2i}{\CI} =  \frac{16 \pi i}{m |\kvec - \kvec'|},
\eeq
and the S-matrix equals $-1$.    The kernel becomes
\beq
\label{kernel}
G(\kvec - \kvec' ) =  \mp  \frac{8 \pi^2}{m |\kvec - \kvec'|} , 
\eeq 
where the $-$ sign corresponds to $g$ being just below the fixed point
$g_*$, where the scattering length $a\to +\infty$ on the BEC side,
whereas the $+$ sign corresponds to $a\to - \infty$ on the BCS side.  
As explained in the Introduction,  we work on the BCS side.

The angular integrals in eq. (\ref{yinteq}) are easily performed.  
Defining the dimensionless variable $\kappa = \kvec^2 / 2mT$,  
the integral equation becomes 
\beq
\label{ykappa}
y (\kappa) =  1 +  4  \int_0^\infty  d\kappa'  \[  
\Theta (\kappa - \kappa') \sqrt{\kappa'/\kappa}  + 
\Theta (\kappa' - \kappa) \]   \frac{z}{e^{\kappa'}  - s z y(\kappa')} 
\eeq
where $z = e^{\mu/T}$ is the fugacity and $\Theta(\kappa)$ is the standard
step function equal to 1 for $\kappa > 0$, zero otherwise.

Finally comparing with the definitions in the last section 
the scaling function for the density  and free energy are
\beq
\label{nhatsc}
q (x)   =  \frac{2}{\sqrt\pi} \int_0^\infty  d\kappa  \sqrt{\kappa} 
  \frac{  y(\kappa) z }{e^\kappa - s  y(\kappa) z}  
\eeq
and 
\beq
\label{cscale}
c =  \frac{2}{\sqrt{\pi} \zeta (5/2) }   
\int_0^\infty  d\kappa \sqrt{\kappa}  \(  -s  \log \( 1- s z y(\kappa)
 e^{-\kappa} \) 
- \inv{2}  \frac{ z ( y(\kappa) - 1 ) }{e^\kappa - s zy(\kappa)} \)
\eeq
The ideal, free  gas limit corresponds to $y=1$ 
where $q= s \Li_{3/2} (s z)$ and  $c= s \Li_{5/2} (sz)/ \zeta(5/2)$,
where $\Li$ is the polylogarithm.    The BEC critical point of the
ideal gas occurs at $\mu=0$, i.e. $q=\zeta(3/2)$.

Consider now two-component fermions with the action (\ref{fermionaction}).  
Here the phase space factor $\CI$ is doubled and since $G\propto 1/\CI$, 
the kernels have an extra $1/2$:
\beq
\label{Gbosferm}
G_{\rm fermi} = \inv{2} G_{\rm bose} 
\eeq
Due to the SU(2) symmetry,  the two-component fermion reduces to
two identical copies of the above 1-component expressions, with the
modification (\ref{Gbosferm}).

\section{Analysis of fermions}

 Recall that for 2 component fermions,
the $4$ is replaced by $2$ in eq. (\ref{ykappa}), and by the SU(2) 
symmetry, the system reduces to two identical copies of the one-component 
equations of the last sections;  below we present results for 
a single  component, it being implicit  that the free energy and density
are doubled.    

The integral equation for $y(\kappa)$, eq. (\ref{ykappa}),  can be
solved numerically by iteration.
 One first substitutes 
$y_0 = 1$ on the right hand side and this gives the  approximation
$y_1$ for $y$.  One then substitutes $y_1$ on the right hand side
to generate $y_2$,  etc.   For regions of $z$ where there are no
critical points,  this procedure converges rapidly, and as little
as 5 iterations are needed.    For fermions,  as one approaches 
zero temperature, i.e. $x$ large and positive,   more iterations are needed
for convergence.  The following results are based on 50 iterations.  

\begin{figure}[htb] 
\begin{center}
\hspace{-15mm} 
\psfrag{x}{$x=\mu/T$}
\psfrag{y}{$c$}
\psfrag{a}{$\rm free$}
\includegraphics[width=10cm]{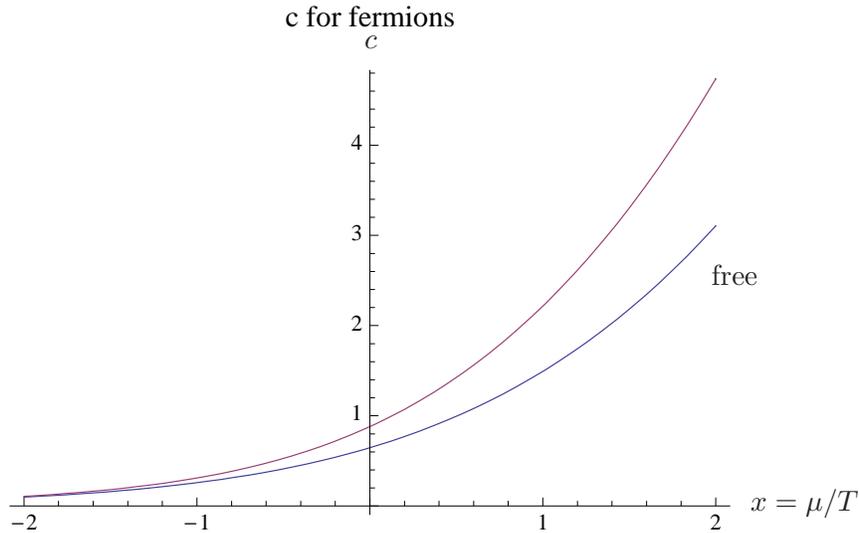} 
\end{center}
\caption{$c(x)$ and its equivalent for a free theory as a function of
$x=\mu/T$.}  
\vspace{-2mm}
\label{cF} 
\end{figure}

When $z\ll 1$,   $y \approx 1$, and the properties of the free ideal 
gas are recovered, since the gas is very dilute.   There are solutions
to eq. (\ref{ykappa}) for all   $-\infty < x < \infty$.  ($x=\mu/T$).  
The scaling function $c$, and it's comparision with a free theory, 
are shown in Figure \ref{cF} as a function of $x$.
   The corrections to the free 
theory become appreciable when $x>-2$.   At $x=0$:
\beq
\label{czero}
c(0) = 0.880, ~~~~~~\xitilde = 0.884, 
\eeq
compared to the free gas values of $c(0) = 0.646$ and $\xitilde =1$.

\begin{figure}[htb] 
\begin{center}
\hspace{-15mm} 
\psfrag{x}{$x=\mu/T$}
\psfrag{y}{$q$}
\psfrag{a}{$\rm free$}
\includegraphics[width=10cm]{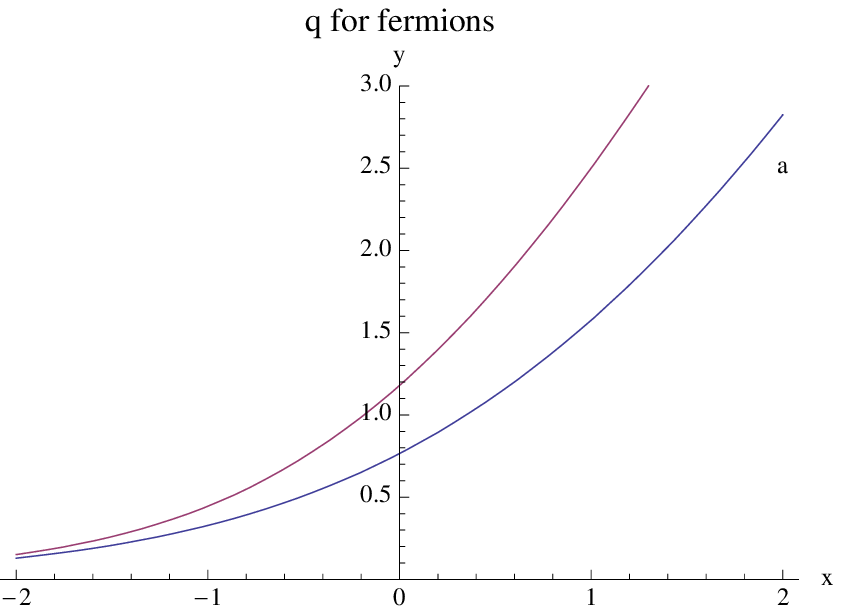} 
\end{center}
\caption{$q(x)$ and its equivalent for a free theory as a function of
$x=\mu/T$.}  
\vspace{-2mm}
\label{qF} 
\end{figure}

\begin{figure}[htb] 
\begin{center}
\hspace{-15mm} 
\psfrag{x}{$\kappa =  \beta \kvec^2 /2m$}
\psfrag{y}{$f$}
\psfrag{a}{$x=15$}
\includegraphics[width=10cm]{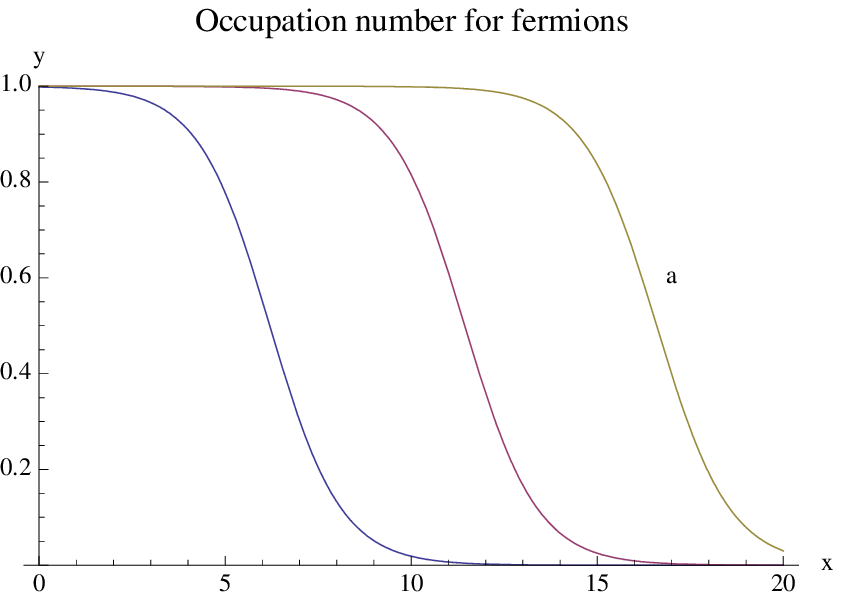} 
\end{center}
\caption{The occupation numbers as a function of $\kappa$ for $x=5,10,15$.}  
\vspace{-2mm}
\label{fF} 
\end{figure}

The scaling function $q$ for the density is shown as function of 
$x$ in Figure \ref{qF}. Note that the density in the interacting case
is always higher than for a free gas,  due to the attractive interactions.
At $x=0$,  $q(0) = 1.18$, whereas for a free gas $q=0.765$.   
At low temperatures and high densities, $\mu/T \gg 1$, the
occupation numbers resemble that of a degenerate Fermi gas,
as shown in Figure \ref{fF}.

\begin{figure}[htb] 
\begin{center}
\hspace{-15mm} 
\psfrag{x}{$x=\mu/T$}
\psfrag{y}{$s/n$}
\includegraphics[width=10cm]{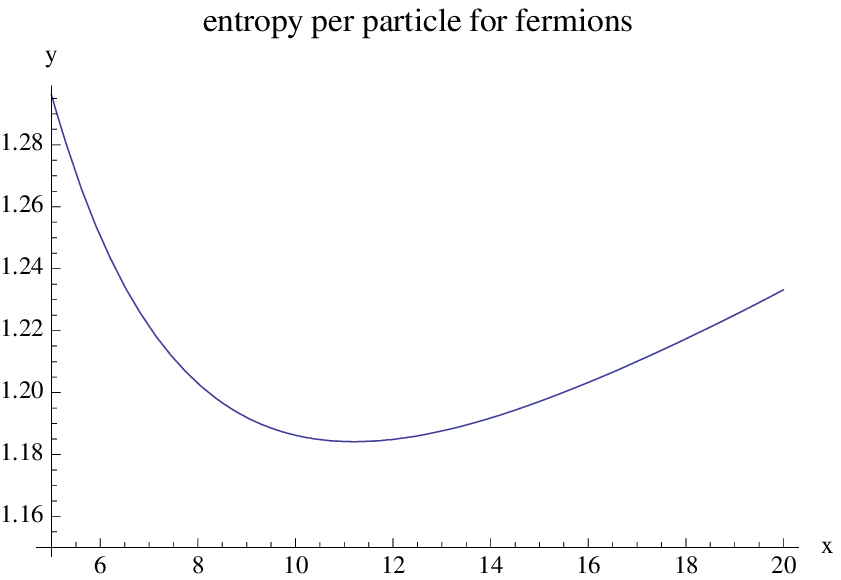} 
\end{center}
\caption{Entropy per fermionic particle as a function of $x$.}  
\vspace{-2mm}
\label{snF} 
\end{figure}

\begin{figure}[htb] 
\begin{center}
\hspace{-15mm} 
\psfrag{x}{$x=\mu/T$}
\psfrag{y}{$C_V/N$}
\includegraphics[width=10cm]{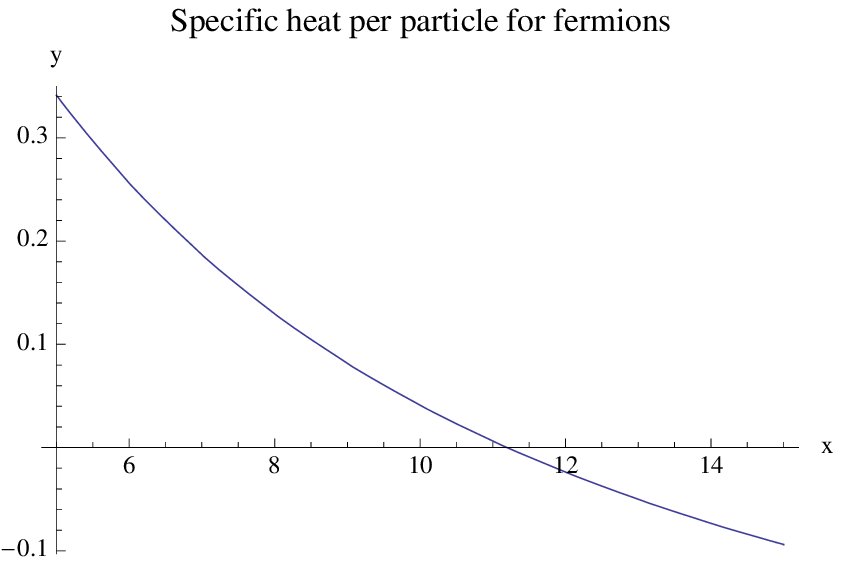} 
\end{center}
\caption{Specific heat per particle as a function of $x$ for fermions.}  
\vspace{-2mm}
\label{CVNF} 
\end{figure}

Whereas $c$ and $q$ are nearly featureless,  other quantities 
seem to indicate a critical point, or phase transition, at 
large density.    For instance,  the entropy per particle 
decreases with decreasing temperature up to $x < x_c \approx 11.2$,
as shown in Figure \ref{snF}.  Beyond this point the entropy per particle
has the unphysical behavior of increasing with temperature.  
A further indication that the region $x>x_c$ is unphysical is
that the specific heat per particle becomes negative, as shown in
Figure \ref{CVNF}.  When $x\ll 0$,  $C_V/N$ approaches the 
classical value $3/2$.   This leads us to suggest a phase transition,
at $x=x_c$, corresponding to the critical temperature  
$ T_c /T_F  \approx 0.1$.  As we will show, our analysis of
the viscosity to entropy-density ratio suggests a higher $T_c/T_F$.  
There have been numerous estimates of $T_c/T_F$ based on various
approximation schemes, mainly using Monte Carlo methods on the lattice
\cite{Perali, LeeShafer,Wingate,Bulgac,Drummond,Burovski}, 
quoting results for $T_c/T_F$ between $0.05$ and $0.23$.  The work
\cite{LeeShafer} puts an upper bound $T_c / T_F < 0.14$,
and the  most recent results of Burovski et. al. quote $T_c/T_F =0.152(7)$.
Our result is thus consistent with previous work.    
The equation of state at this point follows from eq. (\ref{eqnstate}):
\beq
\label{eqstF}
p = 4.95 n T
\eeq

\begin{figure}[htb] 
\begin{center}
\hspace{-15mm} 
\psfrag{X}{$T/T_F$}
\psfrag{Y}{$\frac{E}{N \ep_F}$}
\includegraphics[width=10cm]{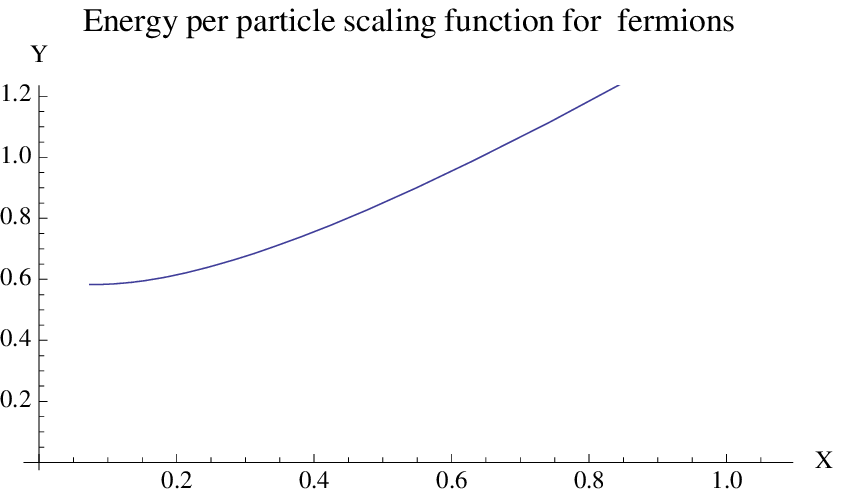} 
\end{center}
\caption{Energy per particle normalized to $\ep_F$  as a function of
$T/T_F$.}  
\vspace{-2mm}
\label{xiF} 
\end{figure}

\begin{figure}[htb] 
\begin{center}
\hspace{-15mm} 
\psfrag{X}{$T/T_F$}
\psfrag{Y}{$s/n$}
\includegraphics[width=10cm]{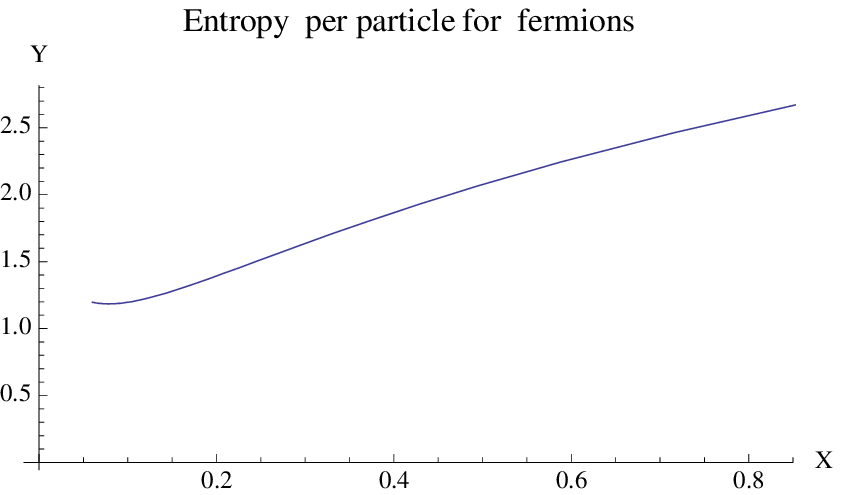} 
\end{center}
\caption{Entopy  per particle   as a function of
$T/T_F$.}  
\vspace{-2mm}
\label{snFTTF} 
\end{figure}

\begin{figure}[htb] 
\begin{center}
\hspace{-15mm}
\psfrag{X}{$T/T_F$}
\psfrag{Y}{$\mu / \ep_F$}
\includegraphics[width=10cm]{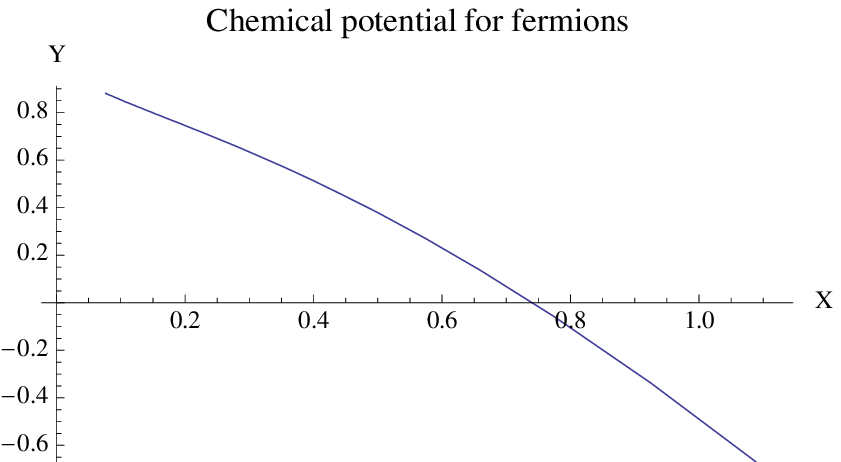} 
\end{center}
\caption{Chemical potential normalized to $\ep_F$ as a function of
$T/T_F$.}  
\vspace{-2mm}
\label{muepF} 
\end{figure}

The energy per particle, normalized to the Fermi energy $\ep_F$,
i.e. $E/N \ep_F = 3 \xi /5$,  and the entropy per particle,  
are  shown in Figures \ref{xiF},\ref{snFTTF} as a function of $T/T_F$, where
$k_B T_F = \ep_F$.   At high temperatures it matches that of a free
Fermi gas, in agreement with the Monte Carlo simulations in
\cite{Bulgac,Burovski}.  Note that there is no sign of 
pair-breaking at $T^*/T_F = 0.5$ predicted in \cite{Perali},
and this also agrees with the Monte Carlo simulations.
 However at low temperatures in the vicinity
of $T_c$, the agreement is not as good.   This suggests our approximation
is breaking down for very large $z$, i.e. the limit of zero temperature. 
The same conclusion is reached by examining $\mu/\ep_F$,
displayed in Figure \ref{muepF}, since the zero temperature 
epsilon expansion and Monte Carlo give
 $\mu/\ep_F \approx 0.4 - 0.5$\cite{Son,Burovski}.

\section{Analysis of bosons}

For bosons we again solved the integral equation (\ref{ykappa}) 
by iteration, starting from $y=1$.   Since the occupation numbers decay
quickly as a function of $\kappa$,  we introduced a cut-off $\kappa < 10$.  
For $x$ less than approximately $-2$,  the gas behaves nearly classically.  

\begin{figure}[htb] 
\begin{center}
\hspace{-15mm}
\psfrag{x}{$x=\mu/T $}
\psfrag{y}{$c$}
\psfrag{a}{$\rm ideal$}
\includegraphics[width=10cm]{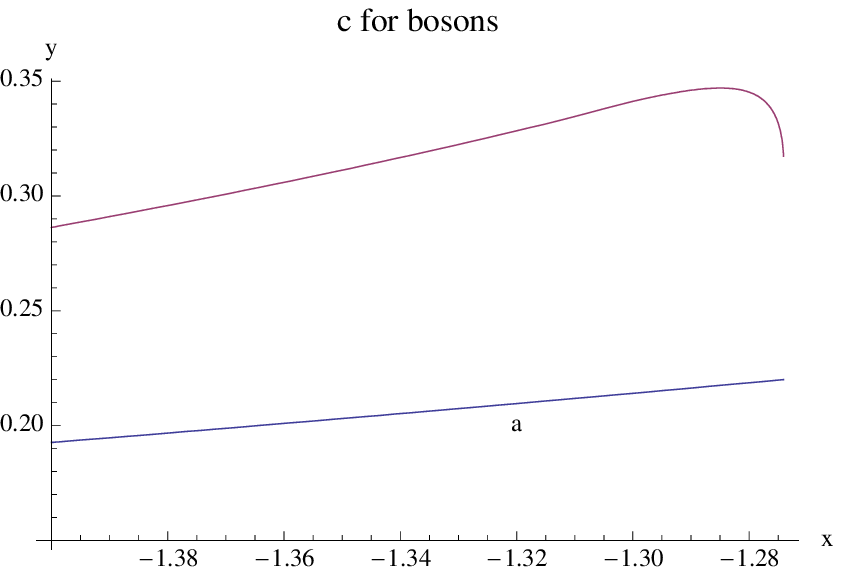} 
\end{center}
\caption{The free-energy scaling function $c$ as a function of $\mu/T$ compared 
to the ideal gas case.}  
\vspace{-2mm}
\label{cB} 
\end{figure}

\begin{figure}[htb] 
\begin{center}
\hspace{-15mm}
\psfrag{x}{$x=\mu/T$}
\psfrag{y}{$\vep (\kvec=0)/T$}
\psfrag{a}{$x_c$}
\includegraphics[width=10cm]{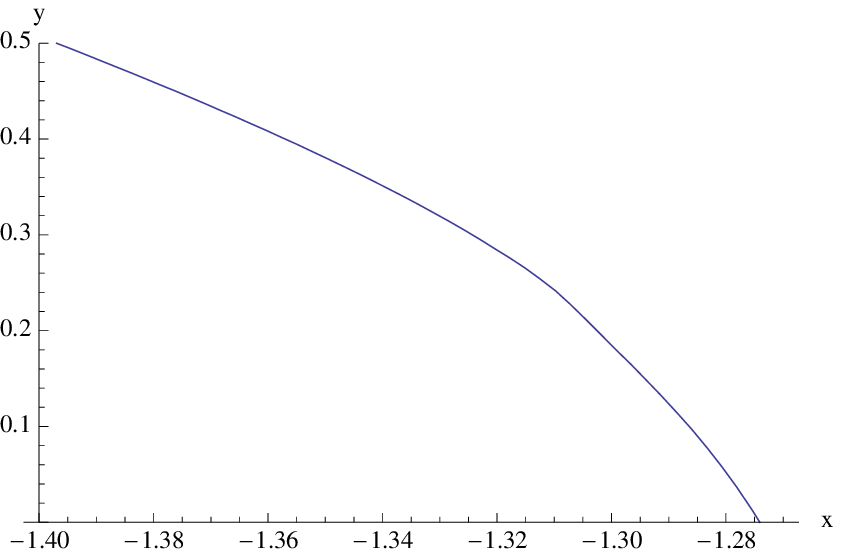} 
\end{center}
\caption{The pseudo-energy $\vep$ at $\kvec =0$  as a function of $x=\mu/T$.}  
\vspace{-2mm}
\label{epofx} 
\end{figure}

\begin{figure}[htb] 
\begin{center}
\hspace{-15mm}
\psfrag{x}{$\kappa =  \beta \kvec^2 /2m $}
\psfrag{y}{$f$}
\psfrag{a}{$x_c$}
\includegraphics[width=10cm]{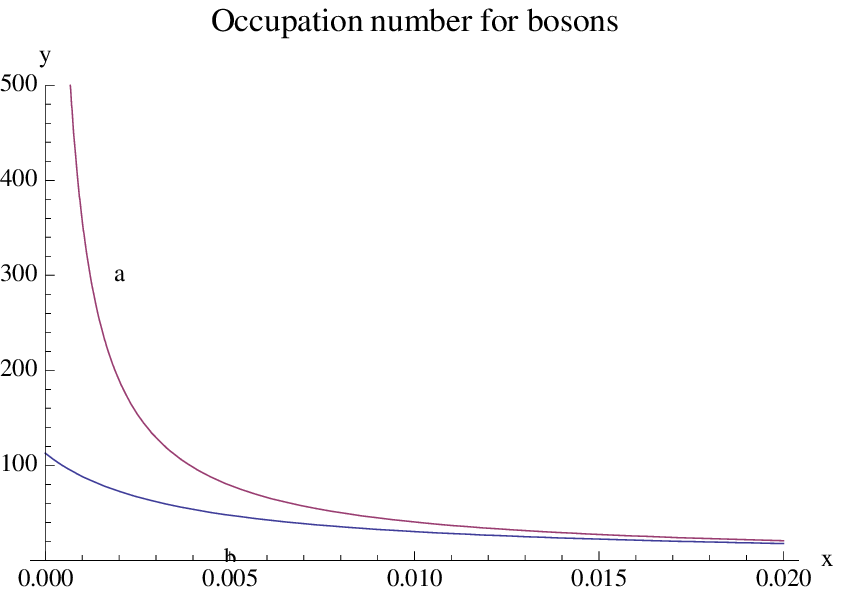} 
\end{center}
\caption{The occupation number $f(\kappa)$  for  $x=-1.275$ and $x_c = -1.2741$.}  
\vspace{-2mm}
\label{fB} 
\end{figure}

The main feature of the solution to the integral equation is that for
$x>x_c \equiv -1.2741$,  there is no solution that is smoothly connected to
the classical limit $x\to -\infty$.  Numerically,  when there is no solution
the iterative procedure fails to converge.  
The free energy scaling function is plotted in Figure \ref{cB}. 
Note that $c<1$, where $c=1$ is the free field value.   
  We thus take the physical region to 
be $x< x_c$.    We find strong evidence that the gas undergoes BEC 
at $x=x_c$.   In Figure \ref{epofx}, we plot $\vep (\kvec=0 )$ as a function of
$x$,  and ones sees that it goes to zero at $x_c$.   This implies the occupation
number $f$ diverges at $\kvec =0$ at this critical point.   
One clearly sees this behavior in Figure \ref{fB}.

\begin{figure}[htb] 
\begin{center}
\hspace{-15mm}
\psfrag{x}{$x=\mu/T $}
\psfrag{y}{$\kappa T (mT)^{3/2}$}
\includegraphics[width=10cm]{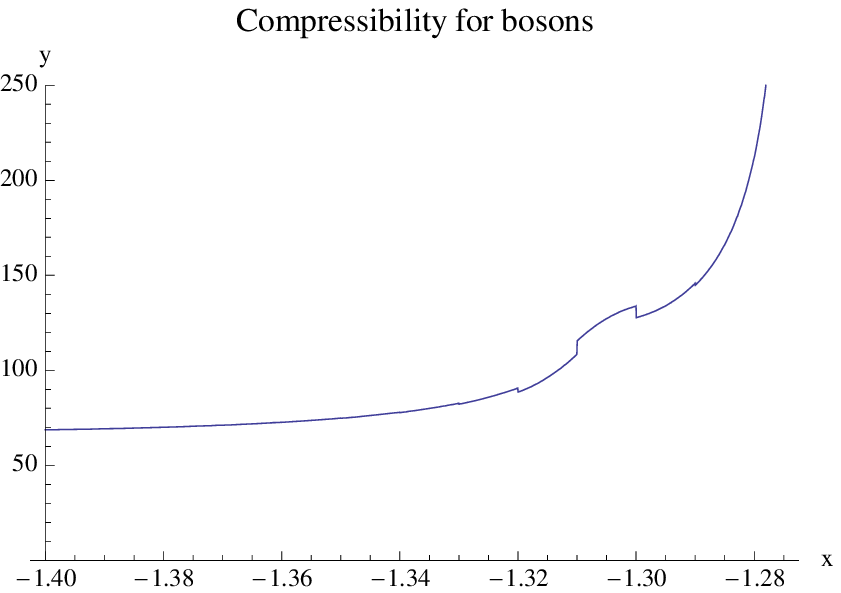} 
\end{center}
\caption{The compressibility  $\kappa $  as a function of $\mu/T$.}  
\vspace{-2mm}
\label{compressB} 
\end{figure}

The compressibility is shown in Figure \ref{compressB},  and diverges at
$x_c$, again consistent with BEC.      
We thus conclude that there is a critical point at $x_c$ which a 
strongly interacting, scale invariant version of the ideal BEC.   
In terms of the density, the critical point is:
\beq
\label{xcb}
n_c \lambda_T^3  =  1.325, ~~~~~~~~~( \mu/T = x_c = -1.2741 )
\eeq
The negative value of the chemical potential is consistent with the
effectively attractive interactions.   The above should be compared with 
the ideal BEC of the free theory,  where $x_c = 0$ and 
$n_c \lambda_T^3 = \zeta (3/2) = 2.61$,  which is higher by a factor of 2.  
At the critical point the equation of state is 
\beq
\label{eqstB}
p = 0.318 n T
\eeq
compared to $p = 0.514 nT$ for the free case.   ($0.514 = \zeta(5/2)/ \zeta (3/2)$). 

A  critical exponent $\nu$ characterizing the diverging  compressibility can be defined as 
\beq
\label{expnu}
\kappa  \sim  (T-T_c)^{-\nu}
\eeq
A log-log plot of the compressibility verses $T-T_c$ shows an approximately
straight line,   and we obtain   $\nu \approx 0.69$.    This should be compared with
BEC in an ideal gas,   where $\nu \approx 1.0$.     Clearly the unitary gas version 
of BEC  is  in a different universality class.

\begin{figure}[htb] 
\begin{center}
\hspace{-15mm}
\psfrag{x}{$T/T_c$}
\psfrag{y}{$s/n$}
\includegraphics[width=10cm]{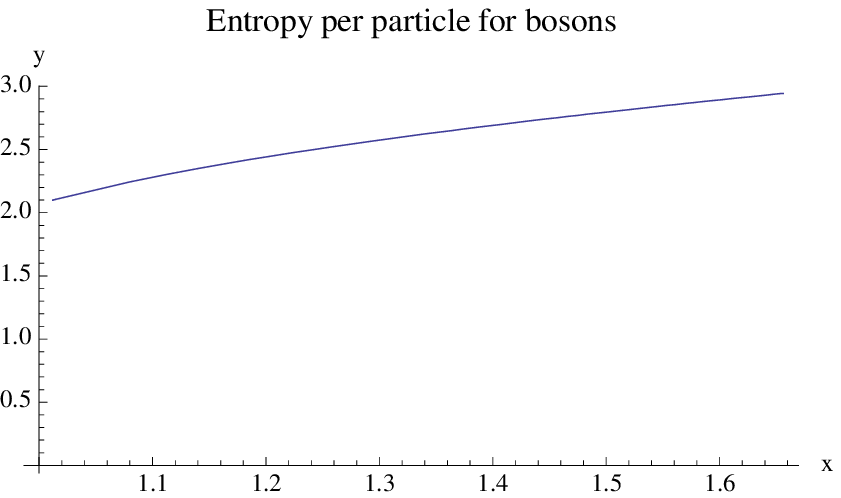} 
\end{center}
\caption{The entropy per particle as a function of  $T/T_c$.}  
\vspace{-2mm}
\label{snB} 
\end{figure}

\begin{figure}[htb] 
\begin{center}
\hspace{-15mm}
\psfrag{x}{$T/T_c$}
\psfrag{y}{$C_V/N $}
\includegraphics[width=10cm]{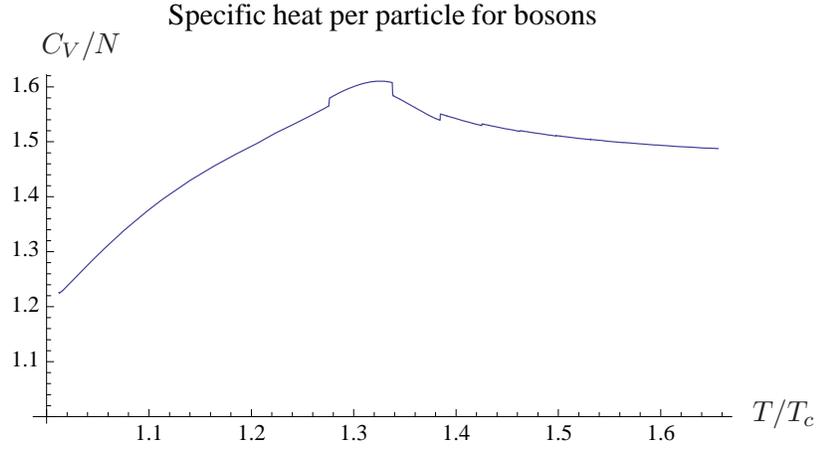} 
\end{center}
\caption{The specific heat per particle as a function of  $T/T_c$.}  
\vspace{-2mm}
\label{CVNB} 
\end{figure}

The energy per particle scaling function $\xitilde$ at the critical point is
$\xitilde (x_c) = 0.281$ compared to $0.453 = (2 \sqrt{2} -2)/(2 \sqrt{2} -1)$ 
for the free case.    The entropy per particle and specific heat per particle
are plotted in Figures \ref{snB}, \ref{CVNB} as a function of $T/T_c$.  
At large temperatures,  as expected $C_V/N = 3/2$, i.e. the classical value. 
It increases as $T$ is lowered,  however in contrast to the ideal gas case,
it then begins to decrease as $T$ approaches $T_c$.

\section{Entropy to viscosity ratio}

Consider first a single component gas.  
In kinetic theory, the shear viscosity can be expressed as 
\beq
\label{shear.1}
\eta = \inv{3}  n  \vbar m \ell_{\rm free} 
\eeq
where $\vbar$ is the average speed and $\ell_{\rm free} $ is the mean
free path.   The mean free path is 
$\ell_{\rm free} = 1/(\sqrt{2} n \sigma)$ where $\sigma$ is the total
cross-section.   (The $\sqrt{2}$  comes from the ratio of the mean speed
to the mean relative speed\cite{Reif}.)     In the unitary limit
the S-matrix $S=-1$,  which implies the scattering amplitude in
eq. (\ref{CMunit}).   This leads to 
\beq
\label{shear.2}
\sigma =  \frac{ m^2 |\CM|^2}{4\pi}  
=  \frac{16 \pi}{|\kvec|^2}
\eeq
where $|k|$ is the momentum of one of the particles in the center of mass
frame,  i.e. $|\kvec_1 - \kvec_2| = 2 |\kvec|$.    This gives
\beq
\label{shear.3}
\eta =  \frac{m^3 \vbar^3}{48 \sqrt{2} \pi} 
\eeq

Since the equation (\ref{energyden}) is  the same relation 
between the pressure and energy of a free gas,  and the pressure
is due to the kinetic energy, this implies
\beq
\label{shear.4}
\inv{2} m \vbar^2  =  E/N  =  \frac{3}{2} \frac{c}{c'} T   
\eeq
Since the  entropy density $s= - \d \CF / \d T$, one finally has
\beq
\label{etas}
\frac{\eta}{s} =  \frac{\sqrt{3\pi}}{8 \zeta (5/2)} 
 \(  \frac{c}{c'} \)^{3/2}  \inv{ 5 c /2 - x c' } 
\eeq

For two-component fermions,  the available phase space $\CI$ is doubled.
Also,  spin up particles only scatter with spin down.  This implies
$\eta$ is $8$ times the above expression.   Since the entropy density is doubled,
this implies that $\eta/s$ is $4$ times the expression eq. (\ref{etas}).

\begin{figure}[htb] 
\begin{center}
\hspace{-15mm} 
\psfrag{X}{$T/T_F$}
\psfrag{Y}{$\eta / s$}
\includegraphics[width=10cm]{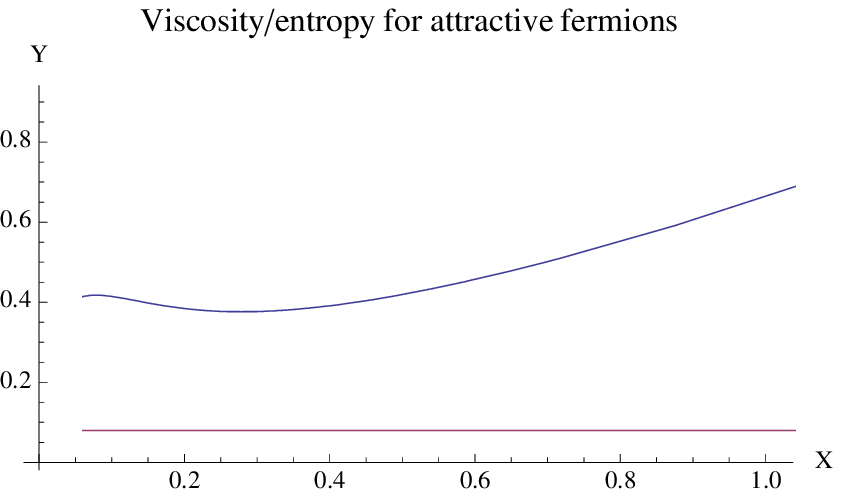} 
\end{center}
\caption{The viscosity to entropy-density ratio  as a function of
$T/T_F$ for fermions.  The horizontal line is $1/4\pi$.}  
\vspace{-2mm}
\label{etasF} 
\end{figure}

The ratio $\eta/s$ for fermions as a function of $T/T_F$ is shown in
Figure \ref{etasF},  and is in good agreement both quantitatively and
qualitatively  with the experimental data 
summarized in \cite{Schafer}.   The lowest value occurs at $x=2.33$, 
which corresponds to $T/T_F = 0.28$, and 
\beq
\label{etasFlim}
\frac{\eta}{s} > 4.72 \frac{\hbar}{4 \pi k_B}
\eeq
The experimental data
has a minimum that is about $6$ times this bound.    
In the free fermion theory the minimum occurs at $\mu/T \approx 2.3$,
which gives $\eta/s >  7.2  \hbar / 4 \pi k_B$.

\begin{figure}[htb] 
\begin{center}
\hspace{-15mm} 
\psfrag{x}{$T/T_c$}
\psfrag{y}{$\eta / s$}
\includegraphics[width=10cm]{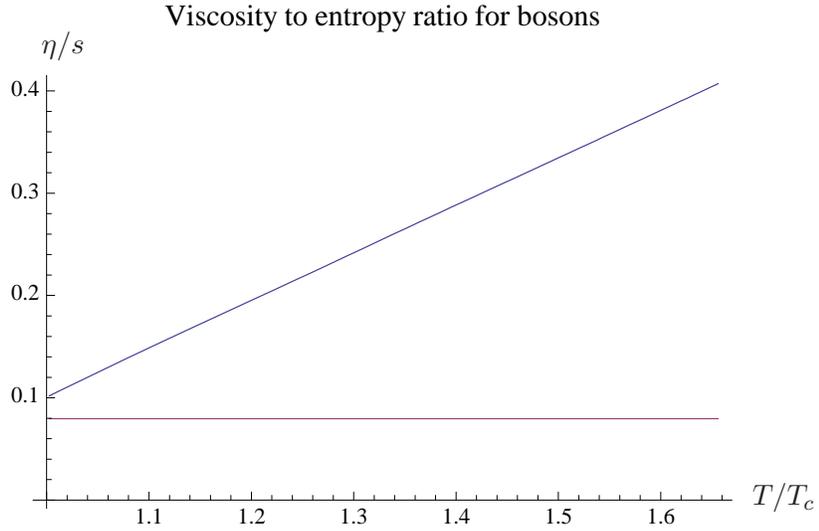} 
\end{center}
\caption{The viscosity to entropy-density ratio  as a function of
$T/T_c$ for bosons.  The horizontal line is $1/4\pi$.}  
\vspace{-2mm}
\label{etasB} 
\end{figure}

For bosons, the ration $\eta/s$ is plotted in Figure \ref{etasB} as
a function of $T/T_c$.   One sees that it has a minimum at the critical point,
where 
\beq
\label{etasBlim}
\frac{\eta}{s} > 1.26  \frac{\hbar}{4 \pi k_B}
\eeq
   Thus the bosonic gas 
at the unitary critical point is a more perfect fluid than that of fermions. 
On the other hand,  the ideal Bose gas at the critical point has a lower value:
\beq
\label{etasBfree}
\frac{\eta}{s} \Bigg\vert_{\rm ideal}  =  \frac{ \sqrt{ 3 \pi \zeta (5/2)}}
{20 \zeta (3/2)^{3/2}} =  0.53 \frac{\hbar}{4 \pi k_B}
\eeq

\section{Conclusions}

We presented a novel analytic treatment of unitary Bose and Fermi gases
at finite temperature and chemical potential 
using  a new formulation of statistical mechanics
based on the exact, zero temperature,  2-body scattering.    
Our results appear to be consistent with lattice Monte Carlo methods. 
All of the thermodynamic functions, such as entropy per particle, energy
per particle,  specific heat, compressibility, and viscosity 
are readily calculated once one numerically solves the integral equation
for the pseudo-energy.    

For fermions, our 2-body approximation is good if the temperatures are not
too low.   We estimated  $T_c/T_F \approx 0.1$,  where the critical 
point occurs at $\mu/T \approx 11.2$.        For bosons we presented
evidence for a strongly interacting version of BEC at the critical 
point $n \lambda_T^3 \approx 1.3$,  corresponding to 
$\mu / T = -1.27$.

\section{Acknowledgments}

We wish to thank Erich Mueller for helpful discussions.    
   This work is supported by the National Science Foundation
under grant number  NSF-PHY-0757868.

\end{document}